# The Science Impact of Astronomy PhD Granting Departments in the United States.

A. L. Kinney, NASA/GSFC, 8800 Greenbelt Road, Greenbelt, MD, 20771

## ABSTRACT

Are you a student applying to Graduate Schools for a PhD in Astronomy? Would you like to know how your favorite departments rank in terms of scientific impact on the field of astronomy? Here, the impact index of the astronomical research of 36 astronomy PhD granting departments is measured and ranked the using the methodology based on Hirsh's h-index and the Molinari and Molinari impact index h(m).

Because of the complex nature of Universities today, this study looks at the Universities in two ways; first analyzing the science impact of the published astrophysical work over a 10 year period of the Department which grants the PhD and; second, looking at the impact of the published astrophysical work from the University as a whole including Laboratories, Centers, and Facilities at the University. The Universities considered in the study are drawn from the 1992 National Research Council study on Programs of Research, Doctorate in Astrophysics and Astronomy (Goldberg, et al., 1995) with three Universities added. Johns Hopkins University, Michigan State University, and Northwestern University all host substantial astronomical research within their Departments of Physics and Astronomy and so are included here.

The first method of measuring impact concentrates on tenured and tenured track faculty, with the top quartile of Universities being 1. Caltech, 2. UC Santa Cruz, 3. Princeton University, 4. Harvard University, 5. University of Colorado, Boulder, 6. SUNY, Stony Brook, 7. Johns Hopkins University, 8. Penn State University, and 9. University of Michigan, Ann Arbor. The second method additionally includes "soft money" scientists in research and adjunct faculty positions, with the top quartile being 1. UC Santa Cruz, 2. Princeton University, 3. Johns Hopkins University, 4. Penn State University, 5. SUNY Stony Brook, 6. University of Michigan, Ann Arbor, 7. New Mexico State University, 8. University of Massachusetts, Amherst, and 9. University of Virginia, Charlottesville.

Both methods reveal important aspects of Universities for graduate students, representing not only the depth but also the breadth of the science available to a student who becomes affiliated with the University.

Finally, a comparison is made of the total articles published in the same 10 year period for the set of universities, both from the departments alone and from the larger universities. Three Universities have both impact index in the top quartile, and additionally have more than 1000 astronomical publications in a decade;

University of California at Santa Cruz, Princeton University, and Johns Hopkins University.

## INTRODUCTION

There is an ever increasing interest in metrics to measure scientific impact of individual scientists (Hirsch, 2005), universities (Molinari & Molinari, 2008), astronomical facilities (Trimble, 2007), and national scientific facilities (Kinney, 2007). The h-index of Hirsch, J.E. (2005) serves as a good independent indicator for evaluating the impact of scientific research. The value of h-index is further enhanced by the fact that it can be meaningfully calibrated for the size of the group being evaluated (Molinari and Molinari, 2008, and Kinney, 2007). The comparisons are particularly useful when made within a single field, where the cultural differences between fields do not come into play. Here, the impact index of the astronomical research of 36 astronomy PhD granting departments is computed, and the institutions are ranked using the methodology based on Hirsh's h-index and the Molinari and Molinari (2008) impact index, h(m).

## METHODOLOGY

Molinari and Molinari (2008) showed that h-index (Hirsch, 2005, the number of publications for which an individual or group has more than that number of citations) of sets of publications produced by groups demonstrate a universal growth rate. Thus h-index can be expressed as the product depending of an index impact, h(m), and a factor that depends on the number of publications, N; h-index = h(m) N^0.4. A subsequent paper by Molinar and Molinari (in press) discusses the mathematical foundation for such behavior. Here we concern ourselves with the fact that this behavior allows a meaningful comparison between the scientific impact of groups of varying size by comparing the impact index; h(m) = h-index/N^0.4.

The Universities considered in this study are drawn from the 1992 National Research Council study on Programs of Research-Doctorate in Astrophysics and Astronomy (Goldberger, et al. 1995) with three universities added. Johns Hopkins University, Michigan State University, and Northwestern University all host substantial astronomical research within their Departments of Physics and Astronomy and so are included here.

The data were compiled from the ISI Web of Knowledge web site (http://isiwebofknowledge.com). Four factors were employed in compiling the total number of publications used in this analysis; PhD granting department affiliation, University affiliation, journal of publication, and year of publication.

### Department affiliation Versus University Affiliation

The PhD granting departments in our sample range from having a rather small number of astronomy faculty to having a large number. However, the more significant differences between the Universities are not size of faculty but rather the fact that some of the Universities have affiliated astronomy and astrophysics Centers and Laboratories, while other Universities do not. For example, the Astronomy Department of Caltech is small, while Caltech has many publishing scientists in affiliated astronomy and astrophysics centers, including Spitzer Science Center, Infrared Processing and Analysis Center, and Jet Propulsion Laboratory. In comparison, the University of California at Santa Cruz has a considerably larger astronomy faculty with affiliation given either as the Department of Astronomy or as Lick Observatory, but UCSC does not have many scientists in their affiliated astronomy or astrophysics Centers or Laboratories.

Because of the varied nature of the Universities analyzed here, the analysis was carried out twice using two different approaches to collecting the publications. First, the publications are collected and analyzed using the department affiliation plus a comprehensive list of astronomical journals and publication year, with results given in Table 1. Second, only the University name, the astronomical journal, and the publication year were used in the search for publications, with results given in Table 2. While the first method provides the very focused view into the scientific impact of the PhD granting department as dominated by the work of tenured and tenure track faculty, the second method gives a sense for the breadth of astronomy affiliated with the University through related departments, centers and laboratories, which help in enhancing research opportunities for PhD students and young scientists at a given University. The additional departments, centers, and laboratories which are actively publishing in astronomy journals are summarized for each University in the Appendix in Table 3. The Appendix also gives the exact commands used in ISI Web of Science to gather the list of total publications.

This approach did not work for Harvard University, where many of the Harvard faculty astronomers are affiliated with the Harvard-Smithsonian Center for Astrophysics, and publish their papers with that affiliation rather than with a Department of Astronomy affiliation. Harvard-Smithsonian CfA has approximately 300 astronomers, most of whom are not faculty of Harvard University. The approach taken here was to compile a list of Harvard faculty astronomers based on the Harvard University Astronomy Department web site, and that list was used to extract the publications list using the same astronomical journals and years of publication as used for the other Universities. The exact commands used are given in the Appendix.

There are challenges when attempting to collect publication from the academic faculty by using the Department name. Some Universities, such as UC Santa Cruz, and UCLA, have changed the names of their Departments recently so that several Department names must be used to collect the correct list of publications. At some Universities, the faculty use various forms of Department names, including departments that do not appear to exist. For example, neither UC Berkeley nor University of Illinois Champagne-Urbana have a Department of Physics and Astronomy, but a number of astronomical publications from those Universities show "Dept Phys Astron" in the address field as seen in ISI Web of Knowledge. Upon further investigation, the scientists in question have joint appointments in the Department of Physics *and* the Department of Astronomy, and most likely the abbreviation "Dept Phys Astron" masks an affiliation given as "Departments of Physics and Astronomy". An inclusive approach has been taken here in order to collect the most complete and correct possible list of publications, using multiple names for degree granting departments when those names appear in active use. The Department names used in the commands in ISI Web of Knowledge to derive the list of publications are given in the last column of Table 1.

Selection by Journal name

There is substantial astronomical research carried out at Universities with no Astronomy Department, either within the Physics Department as in the cases of MIT and Stanford, or within a Physics and Astronomy Department as in the cases of Johns Hopkins University, Michigan State University, and Northwestern University. Because of the desire to look at astronomical research productivity within all of these Departments, the publications analyzed here were selected based on lists of astronomical journals and not based solely on affiliation with a given Department, which in the case of Departments of Physics and Departments of Physics and Astronomy would be dominated by physics publications. Using only the astronomical journals does have the disadvantage of leaving out astronomical articles from publications such as Physical Review, Physical Review Letters, and Science and Nature. While these journals do publish astronomical research, their publications are dominated by physics topics over astronomy by a factor of more than 10 to 1.

The list of astronomical publications was selected so as to collect the greatest and most complete list of publications from several of the leading Astronomy Departments while at the same time excluding journals that have a large percentage of articles which are not related to astronomy.

While it would be preferable to include every astronomy related publication regardless of journal, selection by journal does have additional benefits. Journals themselves have typical impact indices (Molinari and Molinari 2008), which would introduce a source of dispersion in the data. Fields also have their own typical impact indices. Kinney (2007) has shown that the field of physics has a slightly lower impact index than that of astronomy, a further disadvantage to mixing physics publications with astronomy publications. Using the same set of journals gives a good approximation to an unbiased sample of publications upon which to form a comparison between Universities.

Year of Publication

All analysis is done for publications in the years 1993 through 2002. This allows the study to focus on the period after the NRC study of 1992, and yet leaves enough time after the year 2002 so that most of the citations with be achieved.

RESULTS

The data are given in Table 1 for publications produced by astronomy PhD granting departments. The list of 36 departments is ranked by index impact h(m), and the growth curves are given for each ranked quartile in Figures 1 through 4.

Table 1: Impact Index of US Universities: Based on affiliation with University, PhD granting departments for astronomy PhD's.

| Rank | University | N | h-index | h(m) | NRC Rank | Department |
|---|---|---|---|---|---|---|
| 1 | Caltech | 347 | 67 | 6.46 | 1 | Dept Astron |
| 2 | UC Santa Cruz | 1096 | 106 | 6.45 | 6 | (1) |
| 3 | Princeton University | 194 | 51 | 6.20 | 2 | Dept Astron |
| 4 | Harvard University | 757 | 87 | 6.14 | 4 | (2) |
| 5 | Colorado | 256 | 55 | 5.98 | 12 | Dept Astron |
| 6 | SUNY Stony Brook | 209 | 50 | 5.90 | 26 | (3) |
| 7 | JHU (4) | 1112 | 97 | 5.87 | NA | Dept Phys & Astron, Dept Astron |
| 8 | Penn State Univ | 647 | 78 | 5.86 | 21 | Astron & Astrophy, Dept Astron |
| 9 | Univ Michigan | 374 | 62 | 5.79 | 25 | Dept Astron |
| 10 | Univ Hawaii | 995 | 89 | 5.63 | 11 | (5) |
| 11 | Univ Wisconsin | 544 | 70 | 5.63 | 14 | Dept Astron |
| 12 | UC Berkeley | 1210 | 96 | 5.61 | 3 | Dept Astron, Dept Phys & Astron |
| 13 | Michigan State Univ | 196 | 45 | 5.45 | NA | Dept Astron |
| 14 | U Virginia | 474 | 64 | 5.44 | 18 | Dept Astron |
| 15 | New Mexico State U | 316 | 54 | 5.40 | 32 | Dept Astron |
| 16 | MIT | 409 | 59 | 5.32 | 8 | Dept Phys |
| 17 | Yale University | 359 | 56 | 5.32 | 15 | Dept Astron |
| 18 | University of Chicago | 617 | 69 | 5.28 | 5 | Dept Astron |
| 19 | Stanford University | 131 | 37 | 5.26 | 22 | Dept Phys |
| 20 | U Mass Amherst | 297 | 51 | 5.22 | 20 | Dept Astron, Dept Phys & Astron |
| 21 | Ohio State Univ | 751 | 73 | 5.16 | 23 | Dept Astron, Dept Phys & Astron |
| 22 | University of Arizona | 1398 | 93 | 5.13 | 7 | (6) |
| 23 | Univ Texas, Austin | 874 | 76 | 5.06 | 10 | Dept Astron |
| 24 | UCLA | 536 | 62 | 5.02 | 16 | Dept Astron, Dept Phys & Astron, Div Astron |
| 25 | Columbia Univ | 303 | 48 | 4.88 | 18 | Dept Astron |
| 26 | Univ Minnesota | 425 | 55 | 4.88 | 24 | (7) |
| 27 | Univ Florida | 258 | 44 | 4.77 | 31 | Dept Astron |
| 28 | Univ Maryland | 919 | 73 | 4.76 | 19 | Dept Astron, Dept Phys & Astr, Astron Program |
| 29 | Boston Univ | 183 | 37 | 4.60 | 27 | Dept Astron |
| 30 | U Illinois, Ch-Urbana | 535 | 55 | 4.46 | 13 | Dept Astron |
| 31 | Iowa State Univ | 130 | 31 | 4.42 | 30 | Dept Phys & Astron |
| 32 | Univ Indiana | 138 | 31 | 4.31 | 28 | Dept Astron |
| 33 | Northwestern | 216 | 37 | 4.31 | NA | Dept Phys & Astron |
| 34 | Cornell | 419 | 48 | 4.29 | 9 | Dept Astron |
| 35 | Georgia State Univ | 127 | 26 | 3.74 | 33 | Dept Phys & Astron |
| 36 | LSU | 150 | 25 | 3.37 | 29 | Dept Phys & Astron |

(1) The University of California at Santa Cruz formerly used the address "Board Studies of Astronomy and Astrophysics" but now uses "Astronomy Department". In addition, astronomy faculty members at UCSC have as an address "Lick Observatory". The address used in the search for publications as (AD=Univ Calif Santa Cruz SAME (AD=Dept Astron OR AD=Lick Obs* OR AD=Board Studies Astron & Astrophys)).

(2) Harvard faculty members use the address "Center for Astrophysics" in their publications. If the publications are collected using Center for Astrophysics in the address field, it would draw the publications of all the approximately 300 scientists affiliated with CfA, not just Harvard faculty. Instead, the publications were gathered based on the list of faculty members drawn from the current Harvard web site.

(3) Stony Brook astronomers publish under both the address "Astronomy Program" and "Department of Physics and Astronomy." The address used was AD=Stony Brook SAME (AD=Dept Phys* Astr* OR AD=Astron Prog).

(4) The JHU analysis does not include the NASA funded Space Telescope Science Institute (STScI), which is located on the Johns Hopkins campus in Baltimore but operates as a separate national scientific institute. If the publications for STScI were included in the analysis, the resulting impact index would be reduced to 5.18, with 3461 total publications, and the corresponding h-index rising to 135.

(5) University of Hawaii Institute for Astronomy hosts the graduate students for U Hawaii, while the Department of Physics and Astronomy hosts the undergraduates. The address used for the search was AD= Univ Hawaii SAME (AD=Inst* Astron* OR AD=Astr* Inst* OR AD=Dept* Phys* Astr*).

(6) University of Arizona astronomers publish with affiliation Astronomy Department or Steward Observatory. Of the Steward Observatory staff, approximately 25 are academic staff who teach and host graduate students, while approximately 15 are research astronomers who do not teach and cannot host graduate students officially. The address used for the search was AD = Univ Arizona SAME (AD=Dept Astr* OR AD= Steward Obs*).

(7) University of Minnesota has the Department of Astronomy within the School of Physics and Astronomy. The address used in the publication search was AD=Univ Minnesota SAME (AD=Dept Astron OR AD=Sch Phys & Astron).

Caption for Table 1: Table 1 gives the rank, the University Department, the number of papers, N, h-index, and impact index (h(m)=h-index/N* *0.4), the rank as calculated in the NRC (Goldberger et al. 1995). The h-index is calculated for publications for the 10 years 1993 through 2002. The publications are gathered by requiring the address to be from both the relevant University and the relevant Department, for example "AD=Caltech SAME AD= Dept Astron AND PY=1993-2002. The Department names as used in the search are given in the last column of Table 1.

The growth curves are determined in the following manner. H-index is calculated for the total publications, N, for 1993, then for 1993 plus 1994, and so on up to and including data for all ten years from 1993 to 2002. The data shown in the figures are cumulative, demonstrating how citations increase with an increasing body of work. To give a sense for typical scientific growth curves, two lines are overlaid from Kinney (2007) encompassing the range of growth curves for astronomy, physics, chemistry, engineering, mechanical engineering, and mathematics in the United States.

Previous evaluations of impact index involved large groups of publications, with Molinari and Molinari (2008) analyzing the impact for entire universities, and Kinney (2007) analyzing the impact for non-biomedical science for national science facilities. It is important to verify that the growth curves for publications of University Departments, which are intrinsically smaller organizations, behave in the same manner as for the larger groups. For this reason, the growth curves are shown here, along with the data, which were collected from the Thomson Institute for Scientific Information "Web of Knowledge" web site (http://isiwebofknowledge.com).

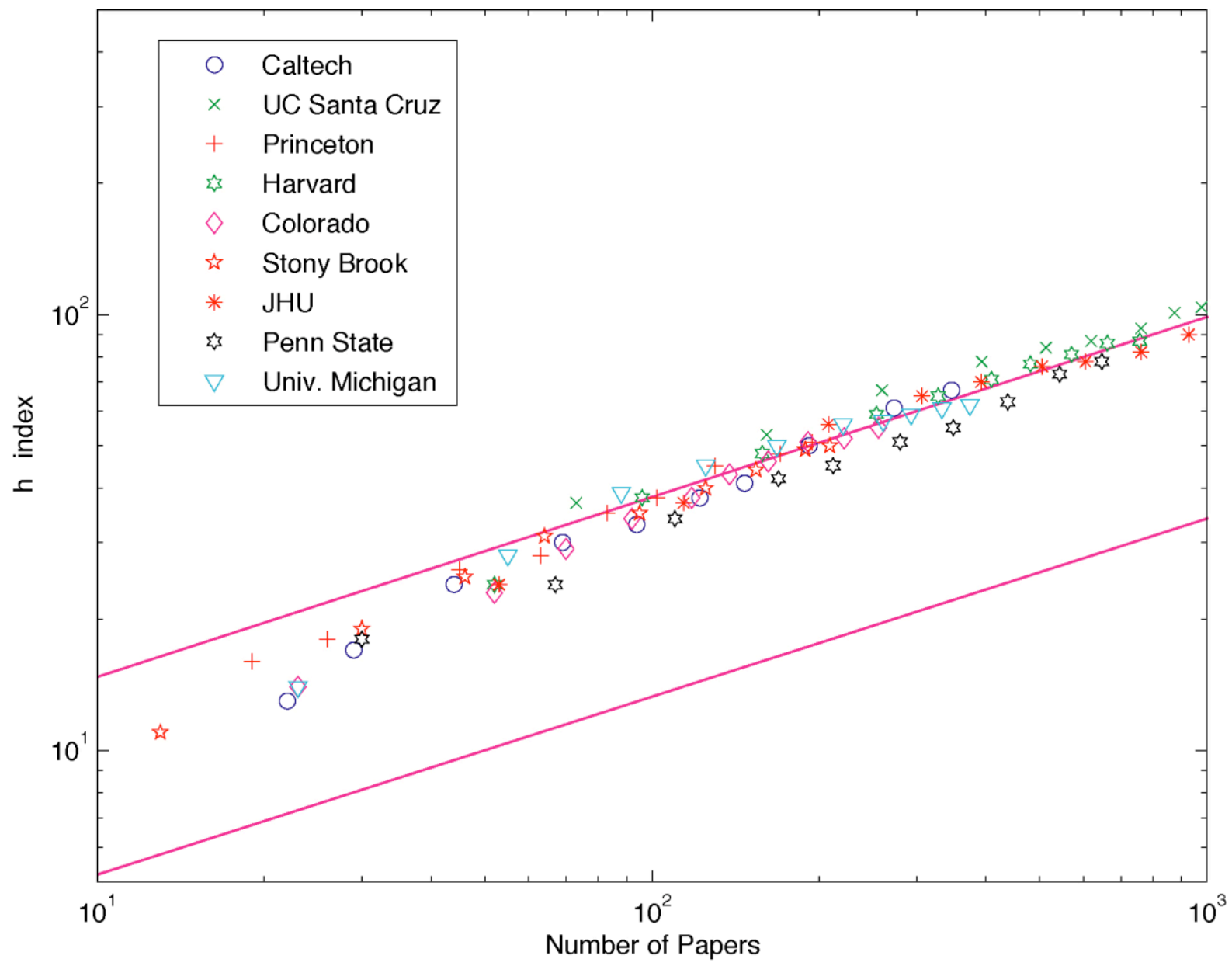

Fig. 1: H-index versus N number of papers for the top quartile of astronomy PhD granting Departments as ranked by h(m). The h-index is calculated for all publications

with authors from the PhD granting department over a ten year span from 1993 to 2002. The data are cumulative, starting with h-index for all publications from 1993, then 1994 plus 1994, and so on, up to and including h-index for all publications from the years 1993 to 2002. Overlaid are lines which show the range in growth curves for non-biomedical science fields including astronomy, physics, chemistry, engineering, mechanical engineering, and mathematics.

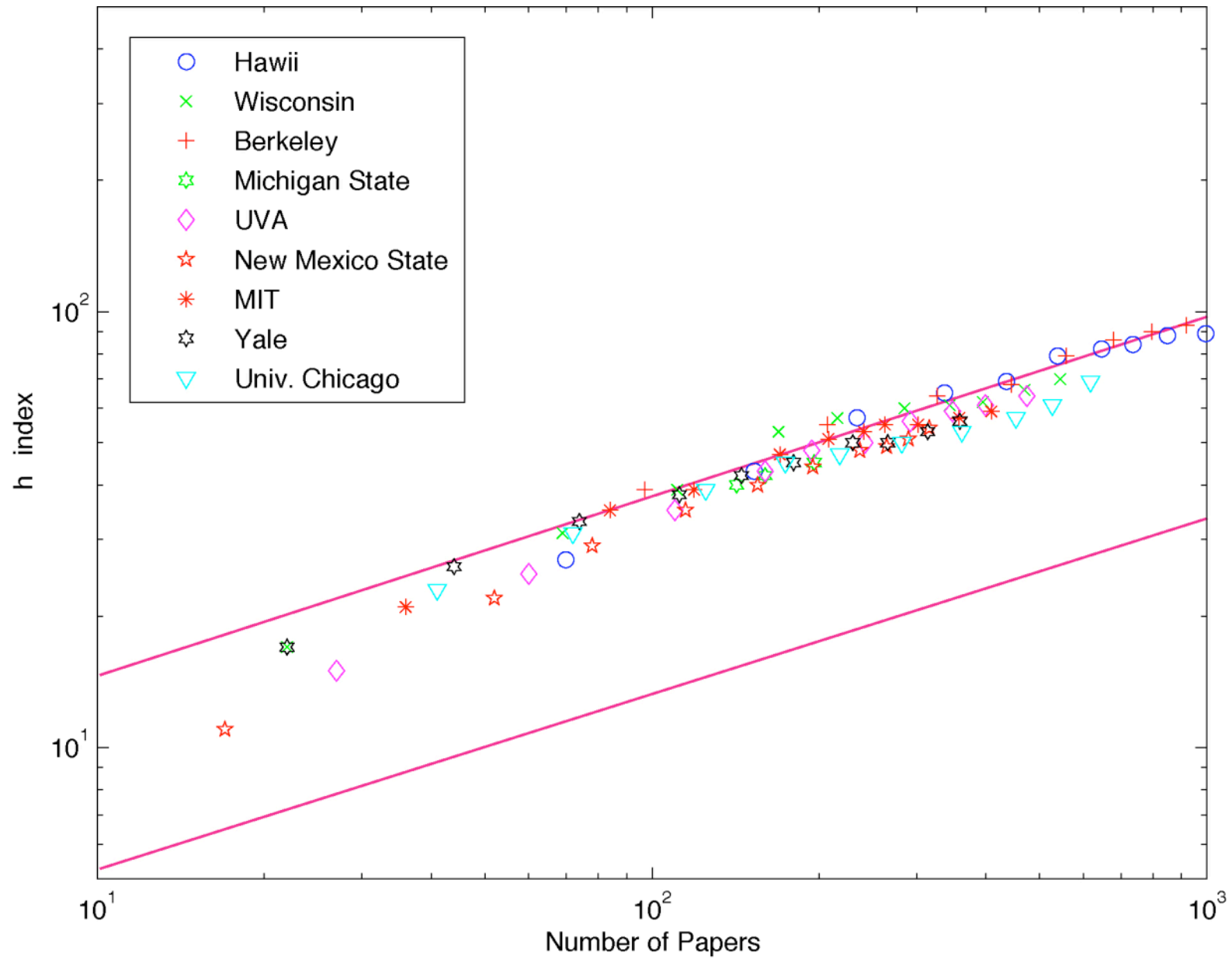

Fig. 2: H-index versus N number of papers for the second quartile of astronomy PhD granting Departments as ranked by impact index h(m).

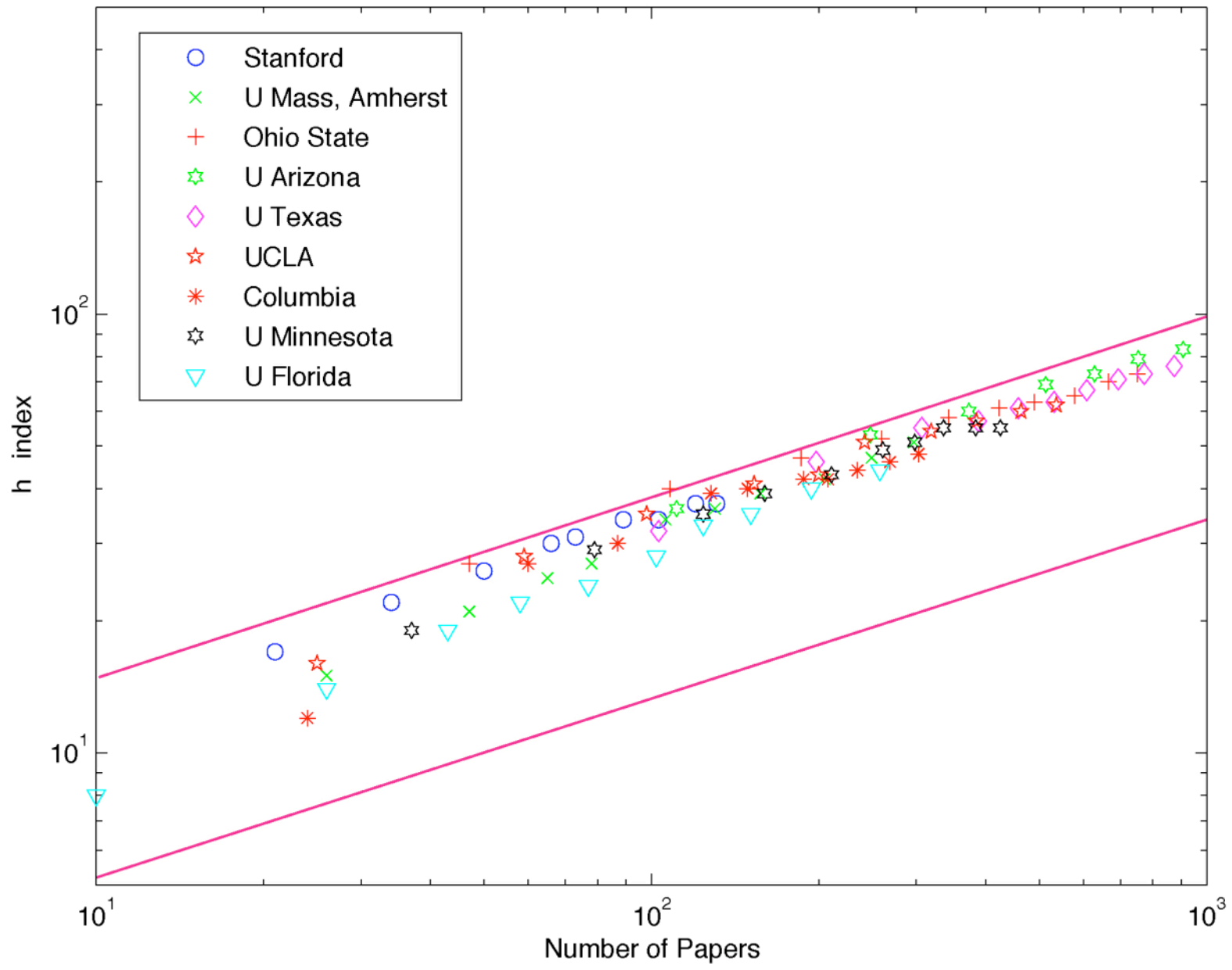

Fig. 3:  H-index versus N number of papers for the third quartile of astronomy PhD granting Departments as ranked by h(m).

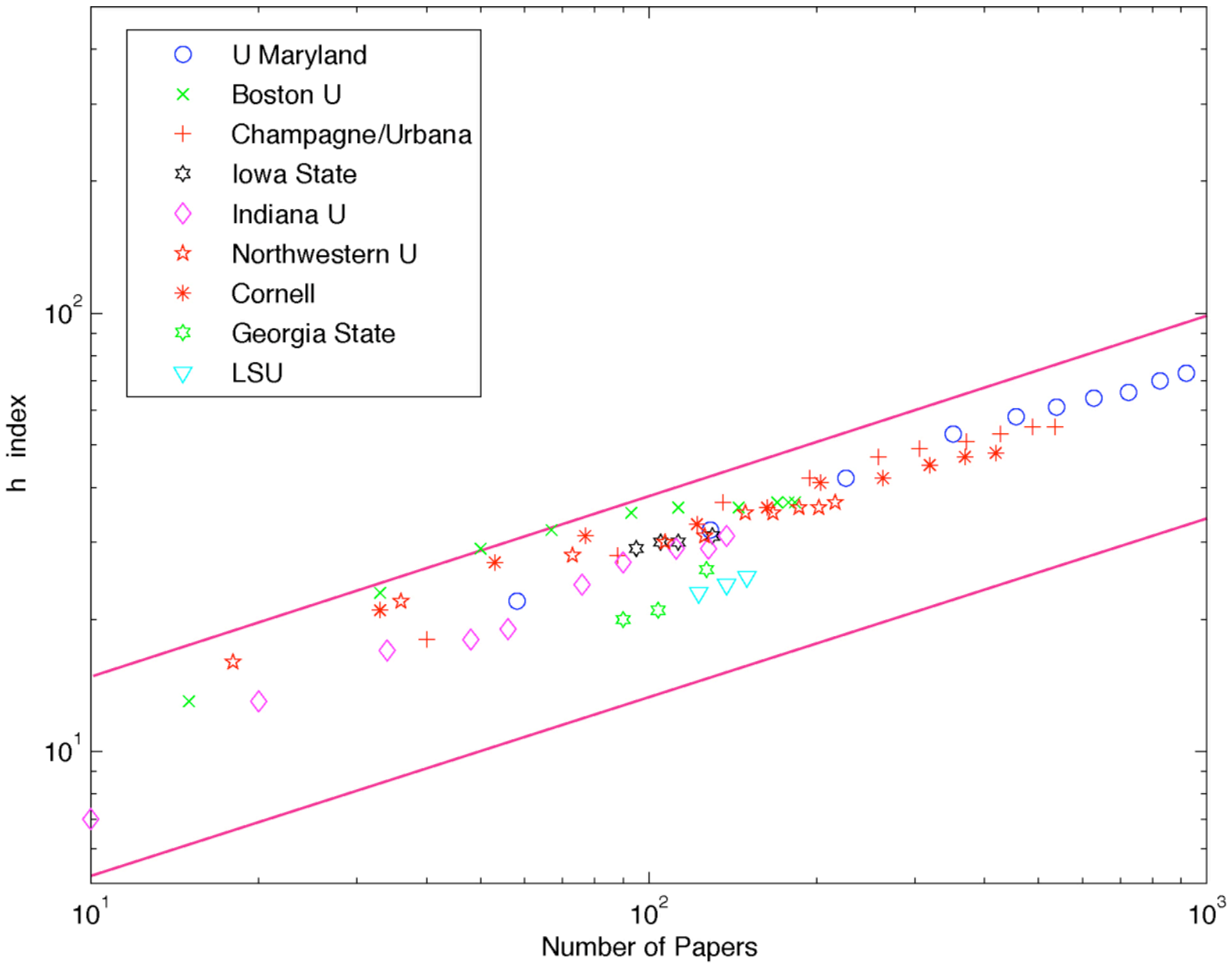

Fig. 4:  H-index versus N number of papers for the forth quartile of astronomy PhD granting Departments as ranked by h(m).

As can be seen in Figures 1 through 4, the universities do indeed show grown curves which follow the standard growth rate of approximately 0.4, so that h-index is well characterized by h(m) N^0.4.   While the numbers of publications are smaller in this study than in previous studies, as noted above, the field of astronomy is a well established field, the journals in which the research is published are well established, and the field is stable.  All these things contribute to a field in which the citations follow the typical behavior of scientific publications, even in a regime of smaller numbers of publications than those previously investigated.  Of the set of fields characterized in Kinney (2007), astronomy was the field with the highest h-index, so it is no surprise that the highest ranking PhD granting departments for astronomy PhD's tend to lie on or above the overlaid lines in the Figures.

The two methods of deriving a publications list from which to measure impact index emphasize two different aspects of a university.   The first method, looking solely at publications from faculty members of the astronomy PhD granting department, yields the highest impact indices, but smaller numbers of publications.   The second method, results show in Table 2, which includes all publications from both faculty members of the astronomy PhD granting department and also affiliated departments, centers, and laboratories results in more publications but lower impact indices.   There are several examples of large differences between the two methods; Caltech, Harvard, and University of Colorado.   In all cases the large difference in impact index traces to having a large number of publishing researchers at affiliated Centers and Laboratories.  In all three cases, the number of publications increases dramatically when looking at the productivity of the Departments plus University affiliates. While the impact index is higher for almost all the Departments in the first method, the large number of astronomy related publications being produced is an important aspect of a Department since it represents the large range of astronomical fields present within the larger community of the University. When choosing a graduate school, the impact of the science produced by the Department is important, but the range of fields available for the student to choose from is also an important factor.

Table 2: Impact Index based on University affiliation.

| Rank | University | N | h-index | h(m) | Table 1 h(m) | Table 1 Rank |
|---|---|---|---|---|---|---|
| 1 | Univ Calif Santa Cruz | 1096 | 110 | 6.40 | 6.45 | 2 |
| 2 | Princeton Univ | 1220 | 104 | 6.06 | 6.20 | 3 |
| 3 | Johns Hopkins Univ (1) | 1587 | 107 | 5.61 | 5.87 | 7 |
| 4 | Penn State Univ | 828 | 82 | 5.58 | 5.86 | 8 |
| 5 | SUNY Stony Brook | 298 | 54 | 5.53 | 5.90 | 6 |
| 6 | Univ Michigan | 861 | 79 | 5.29 | 5.79 | 9 |
| 7 | New Mexico State Univ | 384 | 57 | 5.27 | 5.40 | 15 |
| 8 | U Mass, Amherst | 297 | 51 | 5.23 | 5.36 | 20 |
| 9 | University of Virginia | 586 | 67 | 5.23 | 5.44 | 14 |
| 10 | Michigan State Univ | 218 | 45 | 5.22 | 5.45 | 13 |
| 11 | Yale University | 470 | 61 | 5.20 | 5.32 | 17 |
| 12 | Univ Hawaii | 1259 | 90 | 5.18 | 5.63 | 10 |
| 13 | Univ Wisconsin | 918 | 78 | 5.09 | 5.63 | 11 |
| 14 | University of Chicago | 1103 | 83 | 5.03 | 5.28 | 18 |
| 15 | Ohio State Univ | 846 | 74 | 4.99 | 5.16 | 21 |
| 16 | UC Berkeley | 2504 | 114 | 4.98 | 5.61 | 12 |
| 17 | Harvard | 3510 | 126 | 4.81 | 6.14 | 4 |
| 18 | Univ Minnesota | 504 | 58 | 4.81 | 4.88 | 26 |
| 19 | MIT | 1248 | 83 | 4.79 | 5.32 | 16 |
| 20 | Caltech | 4113 | 107 | 4.76 | 6.46 | 1 |
| 21 | Univ Texas, Austin | 1266 | 83 | 4.76 | 5.06 | 23 |
| 22 | Stanford University | 508 | 57 | 4.71 | 5.26 | 19 |
| 23 | Univ Colorado | 1344 | 84 | 4.70 | 5.98 | 5 |
| 24 | Columbia University | 719 | 64 | 4.61 | 4.88 | 25 |
| 25 | Univ of Illinois, Urbana | 701 | 63 | 4.58* | 4.46 | 30 |
| 26 | Iowa State University | 162 | 32 | 4.56* | 4.42 | 31 |
| 27 | University of Arizona | 2262 | 99 | 4.51 | 5.13 | 22 |
| 28 | University of Florida | 258 | 47 | 4.49 | 4.77 | 27 |
| 29 | University of Maryland | 1260 | 77 | 4.43 | 4.76 | 28 |
| 30 | Boston University | 311 | 43 | 4.33 | 4.60 | 29 |
| 31 | UCLA | 940 | 66 | 4.29 | 5.02 | 24 |
| 32 | Indiana University | 176 | 33 | 4.17 | 4.31 | 32 |
| 33 | Northwestern University | 313 | 41 | 4.12 | 4.31 | 33 |
| 34 | Cornell | 971 | 63 | 4.02 | 4.29 | 34 |
| 35 | Georgia State Univ | 182 | 29 | 3.62 | 3.74 | 36 |
| 36 | LSU | 161 | 25 | 3.27 | 3.37 | 35 |

(1) The analysis for JHU does not include the NASA funded Space Telescope Science Institute (STScI), which is located on the Johns Hopkins campus in Baltimore but operates as a separate national scientific institute. If STScI were to be included in the analysis, the resulting impact index would be reduced to h(m)=5.00, with 3,852 publications and the corresponding h-index rising to 136.

Caption for Table 2: Table 2 gives the rank, number of papers, N, h-index, and impact index (h(m)=h-index/N* *0.4), and the rank as calculated from Table 1. The h-index is calculated for publications for the 10 years 1993 through 2002. The publications are gathered by requiring the address to be from the relevant University regardless of Department.

Note the difference between Caltech and JHU; while both have NASA institutes located on their respective campus, Spitzer Science Center in the case of Caltech, and Space Telescope Science Institute in the case of Johns Hopkins University, the two NASA institutes have different relations with their hosting Universities. Spitzer Science Center is more closely linked to Caltech than STScI is to JHU, with the SSC Director being chosen from amongst current Caltech faculty, and with an overhead paid to the University. While the STScI Director typically has a tenured position at JHU, this position is independently advertised and the Director is not chosen from amongst currently serving JHU faculty members. Also, STScI pays no overhead to JHU. As a reflection of the different relationships, publications from Spitzer Science Center list Caltech in their publication address, while publications from STScI do not list JHU in their publication address.

Given that the volume of publication is also an important indicator for scientific productivity of a university, the number of publications produced during the decade for the 36 Universities is given in histogram form in Figures 5 and 6. It is worth noting that there are a sub-set of schools which seem able to produce a large number of scientific publications while simultaneously maintaining a high science impact. For the PhD granting Universities, both University of California at Santa Cruz and Johns Hopkins University have more than 1000 publications in a 10 year period and fall into the top quartile of impact index (Figure 5). When looking at Universities as a whole, University of California at Santa Cruz, Princeton University and Johns Hopkins University have more than 1000 publications in the 10 year period and fall into the top quartile of impact index, as can be seen in Figure 6.

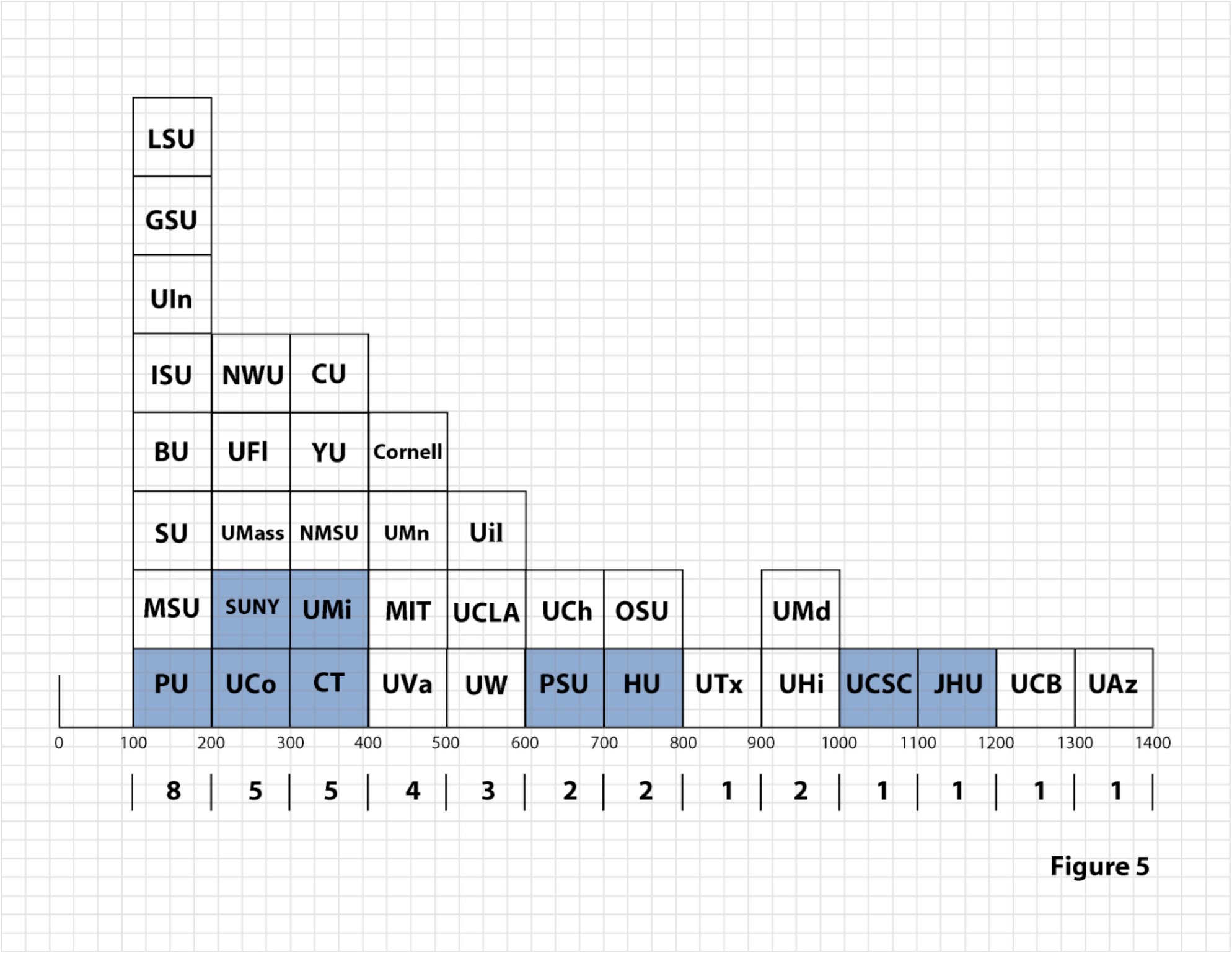

Fig. 5: Histogram showing total number of publications from astronomy PhD granting departments, with those ranked in the top quartile highlighted.

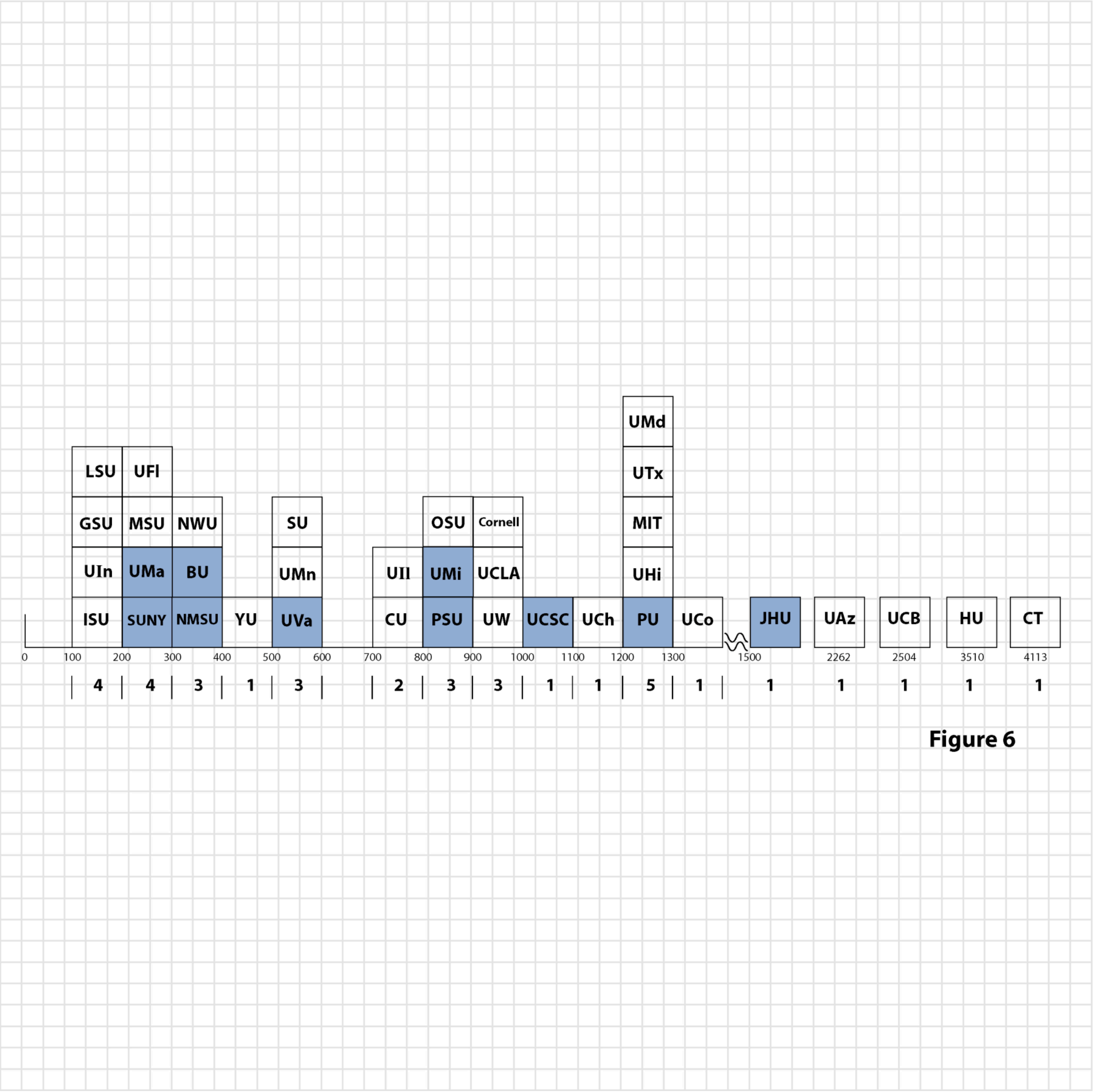

Fig. 6: Histogram showing total number of publications from the 36 Universities covered in this analysis, with those ranked in the top quartile highlighted.

## CONCLUSIONS

H-index and the corresponding impact index, h(m), provide powerful means to evaluate the science impact of the work produced by PhD granting institutions and their affiliated Centers and Laboratories. While the scientific publications of the Departments alone produce higher impact indices, the Departments plus the affiliated Centers and Labs show a higher productivity. Both impact index and productivity are important indicators when choosing a graduate school. A small number of Universities have both a large impact index and produce more than 1000 publications in a decade; University of California at Santa Cruz, Princeton University, and Johns Hopkins University.

APPENDIX

The commands used in ISI Web of Science to derive the h-index and the impact index are given here. The commands included the University name plus city when required, the name or names of the Department, and a list of astronomical and astrophysical journals. The resulting h-indices and impact indices are given in Table 1 of the main paper, and are shown in Figures 1 through 4.

For example, in the first search, the command for Caltech used was:

AD=Caltech SAME AD=Dept Astron AND PY=1993-2002 and (SO=Astronomical Journal OR SO=Monthly Notices of the Royal Astronomical Society OR SO=Astrophysical Journal OR SO=Astronomy & Astrophysics OR SO=Publications of the Astronomical Society of the Pacific OR SO=Astrophysical Journal Supplement Series OR SO=Astronomy & Astrophysics Supplement Series, OR SO=(Astronomy and Space Physics) OR SO=Revista Mexicana de Astronomia Y Astrofisica OR SO=Space Science Reviews OR SO=New Astronomy Reviews OR SO=IAU Symposia OR SO=(Annual Review of Astronomy and Astrophysics) OR SO=Astronomische Nachrichten OR SO=Highlights of Astronomy OR SO=New Astronomy OR SO=Publications of the Astronomical Society of Japan OR SO=(Planetary and Space Science) OR SO=(Astrophysics and Space Science) OR SO=Advances in Space Research OR SO=Icarus OR SO=(Astronomy and Astrophysics) OR SO=Meteoritics & Planetary Science OR SO=(Journal of Geophysical Research Planets) OR SO=Astrophysical Letters & Communications OR SO=Space Astronomy OR SO=Publications of the Astronomical Society of Australia OR SO=(Journal of Astrophysics and Astronomy) OR SO=(General Relativity and Gravitation) OR SO=(Journal of Geophysical Research Space Science) OR SO=Astroparticle Physics OR SO=Experimental Astronomy OR SO=ISIS OR SO=Celestial Mechanics & Dynamical Astronomy OR SO=Journal of the Royal Astronomical Society of Canada OR SO=(Annual Review of Earth and Planetary Sciences) OR SO=Space Communications)

In the case of Harvard University, many of the faculty of the Department of Astronomy publish under the address of "Harvard Smithsonian Center for Astrophysics", while they also sometimes publish under the address "Department of Astronomy". If the CfA address is used in selecting publications, then the publications for the entire staff of CfA are also included. Instead, the names of the Harvard University Astronomy Department faculty, as listed on their web site, was used here to gather their publications. The commands were divided into two because of the ISI Web of Science limit of no more than 50 selection parameters per command. The two selection statements are given here:

AD=Harvard and (au=alcock c* or au=Charbonneau d* or au=dalgarno a* or au=finkbeiner d* or au=gaensler b* or au=grindlay j* or au=hernquist l* or au=huchra j* or au=kirshner r*) and PY=1993-2002 and (SO=Astronomical Journal OR SO=Monthly Notices of the Royal Astronomical Society OR SO=Astrophysical Journal OR SO=Astronomy & Astrophysics OR SO=Publications of the Astronomical Society of the Pacific OR SO=Astrophysical Journal Supplement Series OR SO=Astronomy & Astrophysics Supplement Series, OR SO=(Astronomy and Space Physics) OR SO=Revista Mexicana de Astronomia Y Astrofisica OR SO=Space Science Reviews OR SO=New Astronomy Reviews OR SO=IAU Symposia OR SO=(Annual Review of Astronomy and Astrophysics) OR SO=Astronomische Nachrichten OR SO=Highlights of Astronomy OR SO=New Astronomy OR SO=Publications of the Astronomical Society of Japan OR SO=(Planetary and Space Science) OR SO=(Astrophysics and Space Science) OR SO=Advances in Space Research OR SO=Icarus OR SO=(Astronomy and Astrophysics) OR SO=Meteoritics & Planetary Science OR SO=(Journal of Geophysical Research Planets) OR SO=Astrophysical Letters & Communications OR SO=Space Astronomy OR SO=Publications of the Astronomical Society of Australia OR SO=(Journal of Astrophysics and Astronomy) OR SO=(General Relativity and Gravitation) OR SO=(Journal of Geophysical Research Space Science) OR SO=Astroparticle Physics OR SO=Experimental Astronomy OR SO=ISIS OR SO=Celestial Mechanics & Dynamical Astronomy OR SO=Journal of the Royal Astronomical Society of Canada OR SO=(Annual Review of Earth and Planetary Sciences) OR SO=Space Communications)

AD=Harvard and (AU=loeb a* or AU=moran j* or AU=narayan r* or AU=rybicki g* or AU=sasselov d* or AU=shapiro i* or AU=stubbs c* or AU=thaddeus p* or AU=zaldarriaga m* or AU=goodman a*) and PY=1993-2002 and (SO=Astronomical Journal OR SO=Monthly Notices of the Royal Astronomical Society OR SO=Astrophysical Journal OR SO=Astronomy & Astrophysics OR SO=Publications of the Astronomical Society of the Pacific OR SO=Astrophysical Journal Supplement Series OR SO=Astronomy & Astrophysics Supplement Series, OR SO=(Astronomy and Space Physics) OR SO=Revista Mexicana de Astronomia Y Astrofisica OR SO=Space Science Reviews OR SO=New Astronomy Reviews OR SO=IAU Symposia OR SO=(Annual Review of Astronomy and Astrophysics) OR SO=Astronomische Nachrichten OR SO=Highlights of Astronomy OR SO=New Astronomy OR SO=Publications of the Astronomical Society of Japan OR SO=(Planetary and Space Science) OR SO=(Astrophysics and Space Science) OR SO=Advances in Space Research OR SO=Icarus OR SO=(Astronomy and Astrophysics) OR SO=Meteoritics & Planetary

Science OR SO=(Journal of Geophysical Research Planets) OR SO=Astrophysical Letters & Communications OR SO=Space Astronomy OR SO=Publications of the Astronomical Society of Australia OR SO=(Journal of Astrophysics and Astronomy) OR SO=(General Relativity and Gravitation) OR SO=(Journal of Geophysical Research Space Science) OR SO=Astroparticle Physics OR SO=Experimental Astronomy OR SO=ISIS OR SO=Celestial Mechanics & Dynamical Astronomy OR SO=Journal of the Royal Astronomical Society of Canada OR SO=(Annual Review of Earth and Planetary Sciences) OR SO=Space Communications)

A second evaluation was made using only the University in the address and not the PhD granting department.

Table 3: University Departments, Centers, Institutes, Laboratories with astronomical publications:

| University | Cited Departments, Centers |
|---|---|
| UC Santa Cruz | Dept Physics |
| | Dept Earth Science |
| | Santa Cruz Institute of Particle Physics |
| | Institute of Tectonics |
| | Center for Adaptive Optics |
| | Baskin School of Engineering |
| Caltech | Dept Physics |
| | Infrared Processing and Analysis Center (IPAC) |
| | Palomar Observatory |
| | Spitzer Science Center |
| | Jet Propulsion Laboratory |
| | Division of Geological & Planetary Science |
| Princeton University | Princeton University Observatory |
| | Dept Physics |
| | Joseph Henry Labs |
| Harvard | Harvard-Smithsonian Center for Astrophysics |
| | Dept Physics |
| | Division of Applied Science |
| | Lyman Laboratory of Physics |
| Univ Colorado | Joint Institute of Lab Astrophysics |
| | Center for Astrophysics and Space Astronomy |
| | Laboratory for Atmospheric and Space Physics |
| | National Institute of Standards and Technology |
| | Department of Geological Science |
| UCLA | Dept of Earth and Space Science |
| | Institute of Geophysics and Planetary Physics |
| | Dept Physics |
| Johns Hopkins University | Center for Astrophysical Sciences |
| | Applied Physics Lab |
| SUNY Stony Brook | Dept Physics |
| | Dept of Earth and Space Science |
| | Institute of Terrestrial and Planetary Atmospheres |
| | Marine Science Research Center |
| Penn State Univ | Center of Gravitational Physics and Geometry |
| | Dept Geoscience |
| | Center Earth System Science |
| University of Michigan | Dept Physics |
| | Dept Atmospheric, Oceanographic & Space Science |
| University of Wisconsin | Dept Physics |
| | Washburn Observatory |
| | Center of Space Science & Engineering |
| Univ Hawaii at Manoa | Dept of Geology and Geophysics |
| | Infrared Telescope Facility |
| | Hawaii Institute of Geophysics & Planetology |
| | School of Ocean & Earth Science & Technology |
| | Canada Hawaii France Telescope Corporation |
| | Science & Technology Institute |
| | Division of Planetary Geoscience |
| | Institute of Geophysics |
| UC Berkeley | Center for Particle Astrophysics |
| | Center for EUV Astrophysics |
| | Space Science Lab |
| | Radio Astronomy Lab |
| | Lawrence Berkeley Lab |
| Michigan State Univ | National Superconducting Cyclotron Lab |
| | Dept Geological Science |
| University of Virginia | School of Engineering and Applied Science |
| | Space Research Lab |
| | Dept of Environmental Science |
| | Virginia Institute of Theoretical Astronomy |
| | National Radio Astronomy Observatory |
| New Mexico State Univ | Dept Physics |
| | Particle Astrophysics Lab |
| U Mass, Amherst | Dept Physics |
| | Five College Radio Astronomy Observatory |
| MIT | Center for Space Research |
| | Haystack Observatory |
| | Bates Linear Accelerator Center |
| | Dept Earth Atmosphere & Planetary Science |
| Yale University | Dept Physics |
| | Yale University Observatory |
| | Yale Southern Observatory |
| University of Chicago | Enrico Fermi Institute |
| | Astrophysics & Space Research Lab |
| | Center Astronomy & Astrophysics |
| | Dept Geophysical Science |
| | Yerkes Observatory |
| | Dept Physics |
| Stanford University | Hansen Experimental Physics Lab |
| | Dept Geophysics |
| | Center Space Astronomy & Astrophysics |
| | Dept Applied Physics |
| | Dept Geophysics |
| | Division of Applied Mechanics |
| | Center Radar Astronomy |
| | System Optimization Lab |
| Ohio State University | Dept Physics |
| | Dept Microbiology |
| | Byrd Polar Research Center |
| | Dept Civil and Environmental Engineering |
| University of Arizona | Dept Planetary Science, Lunar & Planetary Lab |
| | Dept Physics |
| | Multiple Mirror Telescope Observatory |
| | Submillimeter Telescope Observatory |
| | Institute of Astronomy |
| Univ Texas, Austin | Center Lithosphere Studies |
| | McDonald Observatory |
| Columbia University | Barnard College |
| | Columbia Astrophysics Lab |
| | Physics Lab |
| | Columbia Computer Lab |
| University of Florida | Dept Physics |
| | Dept Geology |
| | Dept Chemistry |
| | Center Chemical Physics |
| | Rosemary Hill Observatory |
| University of Maryland | Dept Geology |
| | Dept Physics |
| | Laboratory for Millimeter Wave Astronomy |
| Boston University | Microelectronics Research Lab |
| | Institute Astrophysical Research |
| | Center Space Physics |
| | Dept Physics |
| | Dept Mathematics & Statistics |
| Univ Illinois, Urbana | Dept Physics |
| | Center Theoretical Physics |
| | National Center Supercomputing Application |
| | Laboratory Astronomical Imaging |
| Iowa State Univ | Science & Technology |
| | Dept Aerospace Engineering |
| | Erwin W. Fick Observatory |
| | Dept Geology & Atmospheric Science |
| Indiana University | Dept Geology |
| | Dept Physics |
| Northwestern University | Dept Geological Science |
| | Dearborn Observatory |
| Cornell University | Center of Radiophyics & Space Research (CRSR) |
| | National Astronomy & Ionosphere Center (NAIC) |
| | Floyd R. Newman Lab of Nuclear Studies |
| | Dept Geological Science |
| | Applied Physics Lab |
| Georgia State University | Center High Resolution Imaging Astronomy |
| | Dept Physics |
| Louisiana State University | Institute of Environmental Science |
| | Center for Agriculture |
| | Dept Geology & Geophysics |

Publications on the list selected on University plus Department affiliation from Table 1 were subtracted from the publications on the list selected solely on University affiliation. The resulting publications were ranked by the number of citations, and then the top 30 cited papers were reviewed. The Department or Center on these top cited papers are given in this table. Often the publications had no Department or Center affiliation.